\documentstyle[11pt]{article}

\catcode`\@=11
\long\def\@makefntext#1{
\protect\noindent \hbox to 3.2pt {\hskip-.9pt
$^{{\ninerm\@thefnmark}}$\hfil}#1\hfill}                

 \def\@makefnmark{\hbox to 0pt{$^{\@thefnmark}$\hss}}  
 
\def\ps@myheadings{\let\@mkboth\@gobbletwo
\def\@oddhead{\hbox{}
\rightmark\hfil\ninerm\thepage}
\def\@oddfoot{}\def\@evenhead{\ninerm\thepage\hfil
\leftmark\hbox{}}\def\@evenfoot{}
\def\sectionmark##1{}\def\subsectionmark##1{}}
 
\newcounter{sectionc}\newcounter{subsectionc}\newcounter{subsubsectionc}
\renewcommand{\section}[1] {\vspace{0.6cm}\addtocounter{sectionc}{1}
\setcounter{subsectionc}{0}\setcounter{subsubsectionc}{0}\noindent
	{\bf\thesectionc. #1}\par\vspace{0.4cm}}
\renewcommand{\subsection}[1] {\vspace{0.6cm}\addtocounter{subsectionc}{1}
	\setcounter{subsubsectionc}{0}\noindent
	{\it\thesectionc.\thesubsectionc. #1}\par\vspace{0.4cm}}
\renewcommand{\subsubsection}[1] {\vspace{0.6cm}\addtocounter{subsubsectionc}{1}
	\noindent {\rm\thesectionc.\thesubsectionc.\thesubsubsectionc.
	#1}\par\vspace{0.4cm}}

\newcounter{appendixc}
\newcounter{subappendixc}[appendixc]
\newcounter{subsubappendixc}[subappendixc]

\renewcommand{\appendix}[1] {\vspace{0.6cm}
	\refstepcounter{appendixc}
	\setcounter{figure}{0}
	\setcounter{table}{0}
	\setcounter{equation}{0}
	\renewcommand{\thefigure}{\Alph{appendixc}.\arabic{figure}}
	\renewcommand{\thetable}{\Alph{appendixc}.\arabic{table}}
	\renewcommand{\theappendixc}{\Alph{appendixc}}
	\renewcommand{\theequation}{\Alph{appendixc}.\arabic{equation}}
	\noindent{\bf Appendix \theappendixc #1}\par\vspace{0.4cm}}

\renewenvironment{thebibliography}[1]
	{\begin{list}{\arabic{enumi}.}
	{\usecounter{enumi}\setlength{\parsep}{0pt}
\setlength{\leftmargin 1.25cm}{\rightmargin 0pt}
	 \setlength{\itemsep}{0pt} \settowidth
	{\labelwidth}{#1.}\sloppy}}{\end{list}}
 
\newcounter{itemlistc}
\newcounter{romanlistc}
\newcounter{alphlistc}
\newcounter{arabiclistc}

\newcommand{\fcaption}[1]{
	\refstepcounter{figure}
	\setbox\@tempboxa = \hbox{\tenrm Fig.~\thefigure. #1}
	\ifdim \wd\@tempboxa > 6in
	   {\begin{center}
	\parbox{6in}{\tenrm\baselineskip=12pt Fig.~\thefigure. #1}
	    \end{center}}
	\else
	     {\begin{center}
	     {\tenrm Fig.~\thefigure. #1}
	      \end{center}}
	\fi}
 
\newcommand{\tcaption}[1]{
	\refstepcounter{table}
	\setbox\@tempboxa = \hbox{\tenrm Table~\thetable. #1}
	\ifdim \wd\@tempboxa > 6in
	   {\begin{center}
	\parbox{6in}{\tenrm\baselineskip=12pt Table~\thetable. #1}
	    \end{center}}
	\else
	     {\begin{center}
	     {\tenrm Table~\thetable. #1}
	      \end{center}}
	\fi}
 
\def\@citex[#1]#2{\if@filesw\immediate\write\@auxout
	{\string\citation{#2}}\fi
\def\@citea{}\@cite{\@for\@citeb:=#2\do
	{\@citea\def\@citea{,}\@ifundefined
	{b@\@citeb}{{\bf ?}\@warning
	{Citation `\@citeb' on page \thepage \space undefined}}
	{\csname b@\@citeb\endcsname}}}{#1}}
 
\newif\if@cghi
\def\cite{\@cghitrue\@ifnextchar [{\@tempswatrue
	\@citex}{\@tempswafalse\@citex[]}}
\def\citelow{\@cghifalse\@ifnextchar [{\@tempswatrue
	\@citex}{\@tempswafalse\@citex[]}}
\def\@cite#1#2{{$\null^{#1}$\if@tempswa\typeout
	{IJCGA warning: optional citation argument
	ignored: `#2'} \fi}}

\def\fnt#1#2{\footnotetext{\kern-.3em
	{$^{\mbox{\sevenrm #1}}$}{#2}}}
 
 1
 1
 1

\font\tenrm=cmr10

\font\ninerm=cmr9

\begin{document}

\vspace{15 mm}
\begin{center}
\large{{\bf PREDICTION OF A FINITE BARE ELECTRICAL CHARGE FROM QUANTUM GRAVITY}}
\end{center}

\vspace{5 mm}
\hspace{2 cm}  J.L. ROSALES \footnote{E-mail: rosales@phyq1.physik.uni-freiburg.de}

\hspace{2 cm}
     {\em Fakult\"at f\"ur Physik, Universit\"at Freiburg,}

\hspace{2 cm} {\em Hermann-Herder-Strasse 3, D-79104 Freiburg, Germany}

\vspace{10 mm}

\begin{quote}
\begin{center}
				Abstract
\end{center}
	
In the spirit of general relativity, spacetime 
should become curved due to the presence of a particle
of a given mass and charge. We
try to understand this fact in the quantum theory of a thin
shell of matter. It leads to  a generalization of
the potential energy of the shell in the semiclassical
highest dominant order of Planck mass. Rather surprisingly,
the quantization of charge is obtained as a 
consequence of the existence of bound states and
the quantum of bare electrical charge is simply 
$e^2= \hbar c/2$.

\end{quote}
\vspace{2 mm}

Pure particle states could be defined by the the total rest mass and
interaction energies which also include self energies. 
In the classical point electron theory it was not possible, however, to
obtain a finite model such that the total mass arises from the coupling
to the electromagnetic field. This is because the self energy diverges 
linearly with the scale length, $a$, without any possible compensation,
\begin{equation}
mc^2\sim e^{2}/2a \mbox{.}
\end{equation}
Of course,  gravity should enter into more realistic discussions and one would
expect that its negative self energy  would lead to a finite compensation 
for Eq. (1) at some scales. In fact, the
picture of a simple spherical shell of charge can not be mantained but,
instead, one should think on a particle as a 
field that curves spacetime due to its own self energy.
A major purpose of theoretical physics is to make a mathematical
model for a physical problem and it seems
reasonable to start by making simple assumptions.
In the present case, the simplest model is the classical 
dynamics of a  spherical thin shell of mass with  
a given electric or magnetic charge embeded in a four dimensional spacetime.
The problem has known solutions since it was earlier solved by 
Israel\cite{kn:Israel} who showed that
the relevant physical description is 
much better  approximated by a non static situation corresponding
to the dynamics of the radius $a$ of the shell:
\begin{equation}
M=-a\{ (1-2M/a+Q^{2}/a^{2}+\dot{a}^2)^{1/2}-(1+\dot{a}^2)^{1/2} \}\mbox{,}
\end{equation}
where, $M=Gm/c^2$, $Q^2=Ge^2/c^4$ define the 
mass and  the charge  of the thin shell 
and derivatives are taken with respect to the  proper time;
the geometry of the shell corresponds to that  of
a three dimensional spacetime section embeded
in the extended four dimensional one. The normal (timelike) coordinate 
to the shell 
is the Gaussian  time of an exterior Reissner-Nordstrom 
geometry. Thus, Eq. (2) should be equivalent 
to the the full set of Einstein equations and  the picture is rather analogous
to that of the proper motion of a bubble of matter inmersed in its own 
curved spacetime.
One can also write the equation of motion in a simpler, suggestive form
\begin{equation}
M=M(1+\dot{a}^2)^{1/2}+\frac{Q^2-M^2}{2a} \mbox{,}
\end{equation}
which represents the total dynamical energy in terms of self interations
and the kinetic energy.  It may be visualized as the effective
hamiltonian constraint:
\begin{equation}
H(a,\dot{a})\equiv M-M(1+\dot{a}^2)^{1/2}-\frac{Q^2-M^2}{2a}=0 \mbox{,}
\end{equation}
which should more properly be expressed in terms of the canonical
momentum,
\begin{equation}
P\equiv \int \frac{\partial H(a,\dot{a})}{\partial \dot{a}} \frac{d\dot{a}}{\dot{a}} = -M sh^{-1}(\dot{a})+G(a) \mbox{.} 
\end{equation}
$G(a)$ so far an arbitrary function which does not change the classical 
equation of motion. 
Feeding it into the Hamiltonian  constraint we  get,
\begin{equation}
H(a,P)\equiv M-\frac{Q^2-M^2}{2a}-M ch(\frac{P-G(a)}{M})=0 \mbox{.}
\end{equation}
It approximately corresponds to the Hamitonian 
derived by Berezin et al.\cite{kn:Berezin},
however, the ADM and shell masses were treated by Berezin 
as independent
variables; on the other hand, we are here
concerned on the case of charged particles  rather than in
black holes and, correspondingly, our analysis will lead to 
different and independent conclusions. 
Thus, $Q^2>M^2$ (but notice that 
such a condition does not corresponds to the existence of 
naked singulaties because they are absent, by construction, 
in the thin shell model). 

The standard quantization procedure consists on replacing $a$ and $P$ by 
quantum operators satisfying Dirac commutation relation acting on 
the wave functional of the dynamical variable ($L_{Pl}^{2}$ stands for Planck
length squared)
\begin{equation}
[\hat{P},\hat{a}]=-iL_{Pl}^{2} \mbox{,}
\end{equation}
thus, the Hamiltonian constraint leads (modulo factor ordering) to the Wheeler-deWitt equation:
\begin{equation}                            
\{ M-\frac{Q^2-M^2}{2\hat{a}}-M ch(\frac{\hat{P}-G(\hat{a})}{M})\}\Psi=0 \mbox{.}
\end{equation}
According to the principle of correspondence one would expect the quantum
theory to be approximated by semiclassical solutions of the form: 
\begin{equation}
\Psi\sim K(a)\exp[\frac{i\int P(a)da}{L_{Pl}^2}+O(L_{Pl}^2)] \mbox{,}
\end{equation}
where $K(a)$ is a 
convenient Pauli-Van Vleck-Morette prefactor depending on the factor ordering
and $P(a)$ is obtained
from the classical constraint (hereafter we will write low case letters 
$P/L_{Pl}^{2}\rightarrow p/\hbar$ 
when using standard units):
\begin{equation}
p(a)=g(a)+mc[ ch^{-1}(1+\frac{\alpha}{2mc^2a/e^2})] \mbox{,}
\end{equation}
 $\alpha\equiv Gm^2/e^2-1$. Now, Eq. (6)
states that the total energy of the shell vanishes and  it may also be 
contemplated as a zero energy problem of motion within the action of
an effective potential energy defining the 
classical allowed values of the shell radius,
\begin{equation}
p^2\equiv -m V(a) \mbox{,}
\end{equation}
that is,
\begin{equation}
V(a)=-mc^2 [\frac{g(a)}{mc}+ch^{-1}(1+\frac{\alpha}{2mc^2a/e^2})]^2 \mbox{.}
\end{equation}
On the other hand,  $g(a)=0$ when $x=mc^2a/e^2\gg 1$ for in that case one
should recover the usual post-Newtonian expressions 
(we use $ch^{-1}(1+\alpha/2x)\sim (\alpha/x)^{1/2}$ for $x\gg 1$), 
\begin{equation}
V(a) \sim \frac{e^2}{a}-\frac{Gm^2}{a} \mbox{.}
\end{equation}
Notice that the effective potential energy of the shell has been obtained for
the solutions of  Einstein-Maxwell equations and it should be viewed
as an exact classical result for the shell dynamics. 

For $e^2>Gm^2$ (i.e., $\alpha<0$) and  $x=mc^2a/e^2<|\alpha|/4$,
$V(a)$ would become  a complex number unless 
$g(a)=-mc i\pi$ and it should be the right selection because
the effective potential 
should remain real for all values of the radius of the shell.
These sort of results have to be expected 
in the semiclassical analysis. 
It motivates the definition:
\begin{equation}
V(x)=-mc^2[ch^{-1}(\frac{|\alpha|}{2x}-1)]^{2} \sim -mc^2 [\log(\frac{|\alpha|}{x}-2)]^2 \mbox{,}
\end{equation}
for $x< |\alpha|/4$. This is a very surprising  result: 
even neglecting the action of classical gravity (i.e., when we formally 
put $G\rightarrow 0$), 
an atractive region
arises in the semiclassical limit. There is no
alternative classical analogous to 
this phenomenon. Hence, it seems  to allow the complection of 
the old classical program of 
obtaining a finite 
value of the self energy of the electron. 
On the other hand, quantum-mechanically,  severe restrictions
have to be imposed to the (so far arbitrary) parameters of the potential
in order that there were bound states:
\begin{equation}
2\int_{0}^{|\alpha|[(e^2/mc^2)-Gm^2/c^2]/4}p(a)da\geq \hbar/2 \mbox{,}
\end{equation}
the ground state satisfies the equality, that is (using Eq. (11)),
\begin{equation}
2mc\int_{0}^{|\alpha|/4}ch^{-1}(\frac{|\alpha|}{2x}-1) d[\frac{e^2}{mc^2}x]= \frac{\hbar}{2} \mbox{,}
\end{equation}
and,
\begin{equation}
e^2 = Gm^2+\frac{\hbar c}{2} \mbox{.}
\end{equation}
Therefore, in the limit of negligible gravitational self energy, 
the ground state is defined by $e^2= \hbar c/2$.
It represents the quantization of the electric charge. Of course we must
consider $\hbar c/2$ as the quantum of the bare charge.
We conclude that quantum gravity
should induce 
a finite value of the fine structure constant to high energies.

The above result, 
obtained using a very simplified exact solvable model, 
should be correct to the highest order in the semiclassical expansion for
the solution of the full quantum gravity Wheeler-deWitt equation. 
It is a merit
of the semiclassical program of quantization that this result 
takes place. It also amounts to the quantization of the electric 
charge even without the consideration of magnetic monopoles as
suggested in a classical paper of Dirac\cite{kn:Dirac}. 

I wish to thank the Department of Physics of the University of Freiburg for 
his hospitality. This work is supported by a postdoctoral grant 
from the Spanish Ministry of Education and Culture and the
research project C.I.C. y T., PB 94-0194.

\end{document}